\documentclass{article}  

\usepackage{graphicx}

\usepackage{geometry}
\usepackage{txfonts}
\usepackage{amssymb}
\usepackage{amstext}
\usepackage{natbib}

\usepackage{hyperref}
\newcommand{\Kunit}{\,cm$^{2}\cdot$s$^{-1}$}

\newcommand{\fig}[1]{Fig.~\ref{#1}}

\begin{document} 


\Large
\begin{center}\textbf{A subsolar oxygen abundance or a radiative region deep in Jupiter revealed by thermochemical modelling}\end{center}\normalsize

\large\noindent T. Cavali\'e$^{1,2}$, J. Lunine$^3$, and O. Mousis$^4$\normalsize\\
\vspace{0.2cm}

\noindent$^1$Laboratoire d'Astrophysique de Bordeaux, Univ. Bordeaux, CNRS, B18N, all\'ee Geoffroy Saint-Hilaire, 33615 Pessac, France (ORCID: 0000-0002-0649-1192)\\ 
$^2$LESIA, Observatoire de Paris, PSL Research University, CNRS, Sorbonne Universit\'es, UPMC Univ. Paris 06, Univ. Paris Diderot, Sorbonne Paris Cit\'e, Meudon, France\\
$^3$Cornell University, Ithaca, NY, USA (ORCID: 0000-0003-2279-4131)\\
$^4$Aix Marseille Universit\'e, Institut Origines, CNRS, CNES, LAM, Marseille, France \\

\vspace{0.2cm}
\noindent\textbf{received:} 20 May 2022\\
\noindent\textbf{accepted:} 23 February 2023\\
\vspace{0.2cm}

\noindent\textbf{DOI:} https://doi.org/10.1038/s41550-023-01928-8\\
\vspace{0.5cm}

\section*{Abstract}
Jupiter's deep abundances help to constrain the formation history of the planet and the environment of the protoplanetary nebula. Juno recently measured Jupiter's deep oxygen abundance near the equator to be 2.2$_{-2.1}^{+3.9}$ times the protosolar value (2$\sigma$ uncertainties). Even if the nominal value is supersolar, subsolar abundances cannot be ruled out. Here we use a state-of-the-art one-dimensional thermochemical and diffusion model with updated chemistry to constrain the deep oxygen abundance with upper tropospheric CO observations. We find a value of 0.3$_{-0.2}^{+0.5}$ times the protosolar value. This result suggests that Jupiter could have a carbon-rich envelope that accreted in a region where the protosolar nebula was depleted in water. However, our model can also reproduce a solar/supersolar water abundance if vertical mixing is reduced in a radiative layer where the deep oxygen abundance is obtained. More precise measurements of the deep water abundance are needed to discriminate between these two scenarios and understand Jupiter's internal structure and evolution.

\section{Introduction}
Giant planets are the true architects of planetary systems given their relatively short formation timescales compared with terrestrial planets, their gravity and migration. Their deep composition holds part of the key to understanding the formation of giant planets in planetary systems, along with other measurements such as their gravity and magnetic field. It can also help to constrain the processes that led to the condensation or trapping of the primordial ices in the protosolar nebula \citep{Helled2014,Bar-Nun1988}. 

The Galileo probe plunged in the atmosphere of Jupiter in 1995, reaching the 22 bar level, and measured its elemental and isotopic composition. The main result is the relatively uniform enrichment in volatiles by a factor of two to four with respect to the protosolar value \citep{Wong2004}. The only element that did not follow this trend, and for which there was no internal process to explain its depletion, is oxygen. The accepted hypothesis is that Galileo entered a dry area where water, the main carrier of oxygen at depths, was still not uniformly mixed at 22 bar. The oxygen measurement from Galileo has thus been mostly accepted as a lower limit ever since.

One of the reasons the Juno mission was designed was to fill this gap left by Galileo regarding the deep oxygen abundance in Jupiter. The idea was to observe the microwave spectrum of Jupiter at various emission angles to retrieve the global water abundance\citep{Janssen2005}. The main difficulty then resides in the dependency of the spectrum on the temperature profile and on absorbers other than water, such as ammonia. While Juno confirmed previous observations that ammonia was depleted below its condensation level \citep{dePater1986}, it surprisingly showed that this depletion was latitude dependent and that it extended as deep as $\sim$50 bar \citep{Li2017}, except near the equator. \citet{Li2020} used the microwave spectra seen by Juno at this latitude to retrieve the vertical water profile of Jupiter at its equator and derive its deep oxygen abundance. They found that the oxygen was nominally enriched by a factor of 2.7$^{+2.4}_{-1.7}$ times the protosolar value but the 1$\sigma$ lower bound was weakly determined, and they state that subsolar values are possible. \citet{Helled2022} emphasized that these error bars are sufficiently large that the depletion seen by Galileo may not be a local anomaly, but could instead reflect a global depletion. In what follows, we use protosolar abundances determined by \citet{Lodders2021}. The Jovian oxygen abundance of \citet{Li2020} then translates into 2.2$^{+2.0}_{-1.4}$ times the protosolar value.

Long before Galileo and Juno, the reported \citep{Beer1975} detection of CO in the troposphere of Jupiter with an abundance higher than thermochemical equilibrium predictions by orders of magnitude triggered the development of numerous models to constrain the deep oxygen abundance by solving the balance between thermochemistry and vertical mixing (for example, \citealt{Lunine1987,Fegley1988,Yung1988}). Below the cloud level, in the deep troposphere, CO and H$_2$O are in thermochemical equilibrium resulting from the equilibrium reaction CO $+$ 3H$_2$ $=$ H$_2$O $+$ CH$_4$. As the temperature decreases with height, the thermochemical equilibrium is shifted in favour of H$_2$O. The detection of CO in the upper troposphere then demonstrates that thermochemical equilibrium is quenched by vertical mixing when the mixing timescale becomes shorter than the chemical conversion timescale. Although the initial studies cited above focused on identifying the rate-limiting reaction, more recent work incorporated comprehensive chemical schemes \citep{Visscher2010,Wang2016}. By probing the abundance of CO in the upper troposphere and modelling thermochemistry and diffusion, one can then reconstruct the vertical profile of the species and tie it back to the deep water (and thus to the deep oxygen) abundance in the planet.

We used a 1D thermochemical and diffusion model \citep{Cavalie2017} to fit a CO upper tropospheric mole fraction of 0.9 $\pm$ 0.3 ppb, as measured by previous studies \citep{Bezard2002,Bjoraker2018}. Although the chemical scheme used by \citet{Cavalie2017} was validated globally by the combustion industry over a wide range of temperatures and pressures, \citet{Wang2016} showed that the model failed to agree with most competing chemical schemes regarding the quench chemistry of CO in giant planets. \citet{Moses2014} identified the main source of the disagreement as the kinetics of the conversion reaction of methanol into the methyl radical and water measured by \citet{Hidaka1989}. This led \citet{Venot2020} to fully update their CH$_3$OH submechanism by adopting the work of \citet{Burke2016}, which noticeably provided an explicit logarithmic dependence on pressure of some reaction rates, and removing of the controversial reaction of \citet{Hidaka1989} that produced the methyl radical from methanol (and thus from CO). This updated scheme21 then destroys less CO than that of \citet{Venot2012}, in agreement with other studies \citep{Wang2015,Wang2016,Moses2014} (Extended Data Fig. 1). As a consequence, a given CO abundance requires a lower deep oxygen abundance with the new scheme (Methods).

\section{Results}
\begin{figure*}[!h]
\begin{center}
   \includegraphics[width=15cm,keepaspectratio]{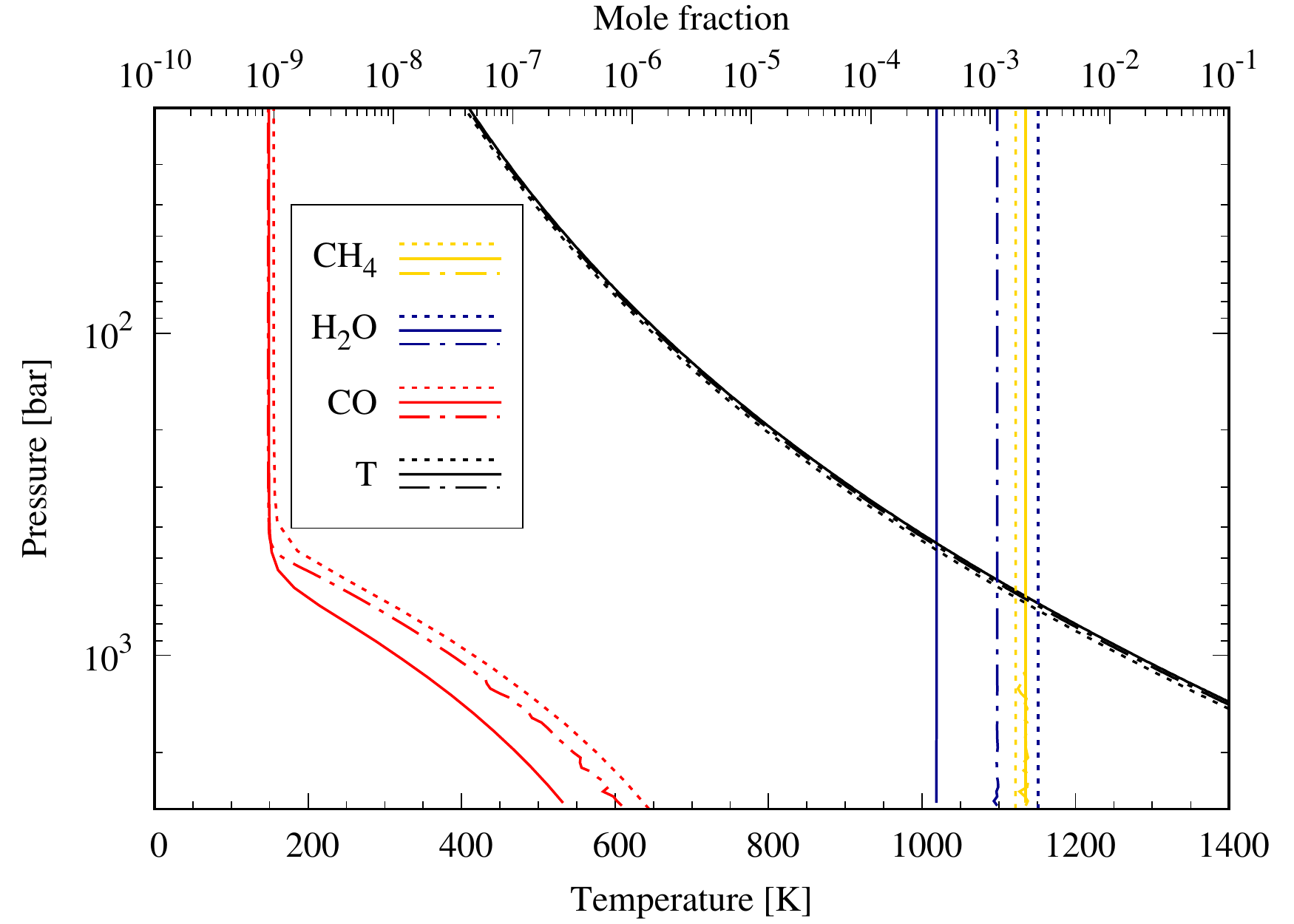}
\end{center}
\caption{Abundances and temperature profiles for Jupiter. Vertical temperature (T), CO, H$_2$O and CH$_4$ profiles from our nominal model with $K_\mathrm{zz}$ $=$ 10$^8$\Kunit~and $y_\mathrm{CH_4}^\mathrm{top}$ $=$ 0.00204 in which oxygen is subsolar (O/H $=$ 0.3 times the protosolar value). The dashed lines correspond to alternative Jupiter abundance and $T$ profiles obtained with a more sluggish mixing ($K_\mathrm{zz}$ is reduced to 2.5 $\times$ 10$^6$\Kunit) to match the nominal Juno O/H value (2.2 times the protosolar value). This lower $K_\mathrm{zz}$ is then indicative of a radiative region located around the quench level of CO (that is, at pressure $p$ $\sim$ 0.4--0.5 kbar and $T$ $\sim$ 1000 K). The dash-dotted lines correspond to a final model in which a deep radiative layer with reduced $K_\mathrm{zz}$ is inserted to match solar oxygen and produce 0.9 ppb CO in Jupiter's upper troposphere: $K_\mathrm{zz}$ is set to 1\Kunit~between $T$ $=$ 1400 K and 2200 K and transitions linearly with $T$ towards our nominal value of 10$^8$\Kunit~at 970 K to ensure that PH$_3$ and GeH$_4$ are quenched at $\sim$800 K. This is a non-unique solution.}
\label{Fig1} 
\end{figure*}

\begin{figure*}[!h]
\begin{center}
   \includegraphics[width=15cm,keepaspectratio]{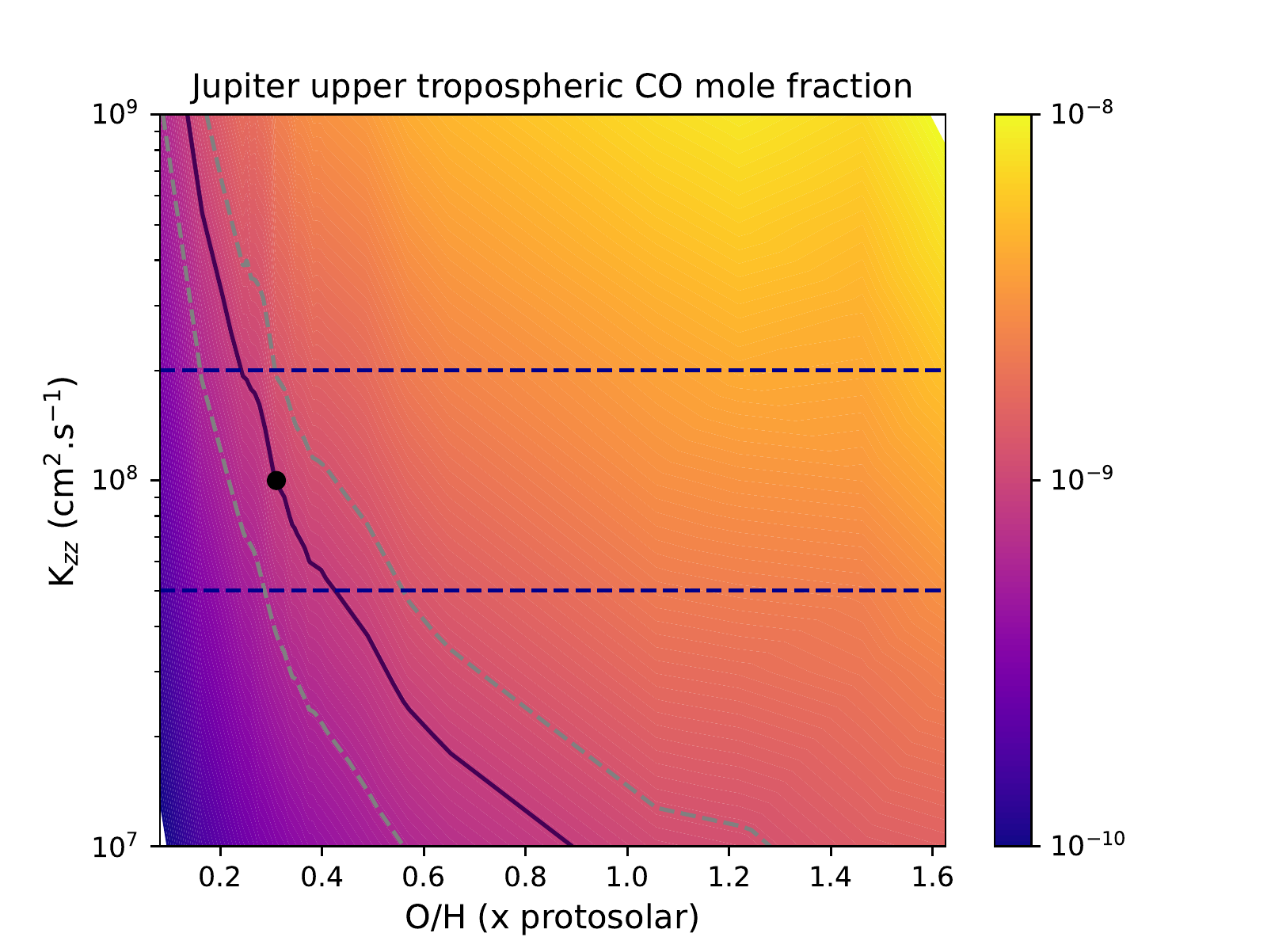}
\end{center}
\caption{$K_\mathrm{zz}$ and oxygen dependence of Jupiter's upper tropospheric CO mole fraction. The CO mole fraction (colour scale) as a function of tropospheric mixing and the deep oxygen abundance relative to solar. The solid line shows the computations that result in 0.9 ppb CO. The grey dashed lines represent the range that results in 0.6 to 1.2 ppb CO (full uncertainty range). The range of possible $K_\mathrm{zz}$ values, constrained from laboratory experiments and matching observed values of PH$_3$ and GeH$_4$, is shown by the dashed blue lines. The nominal value of our model is shown by the black dot.}
\label{Fig2} 
\end{figure*}

\begin{figure*}[!h]
\begin{center}
   \includegraphics[width=15cm,keepaspectratio]{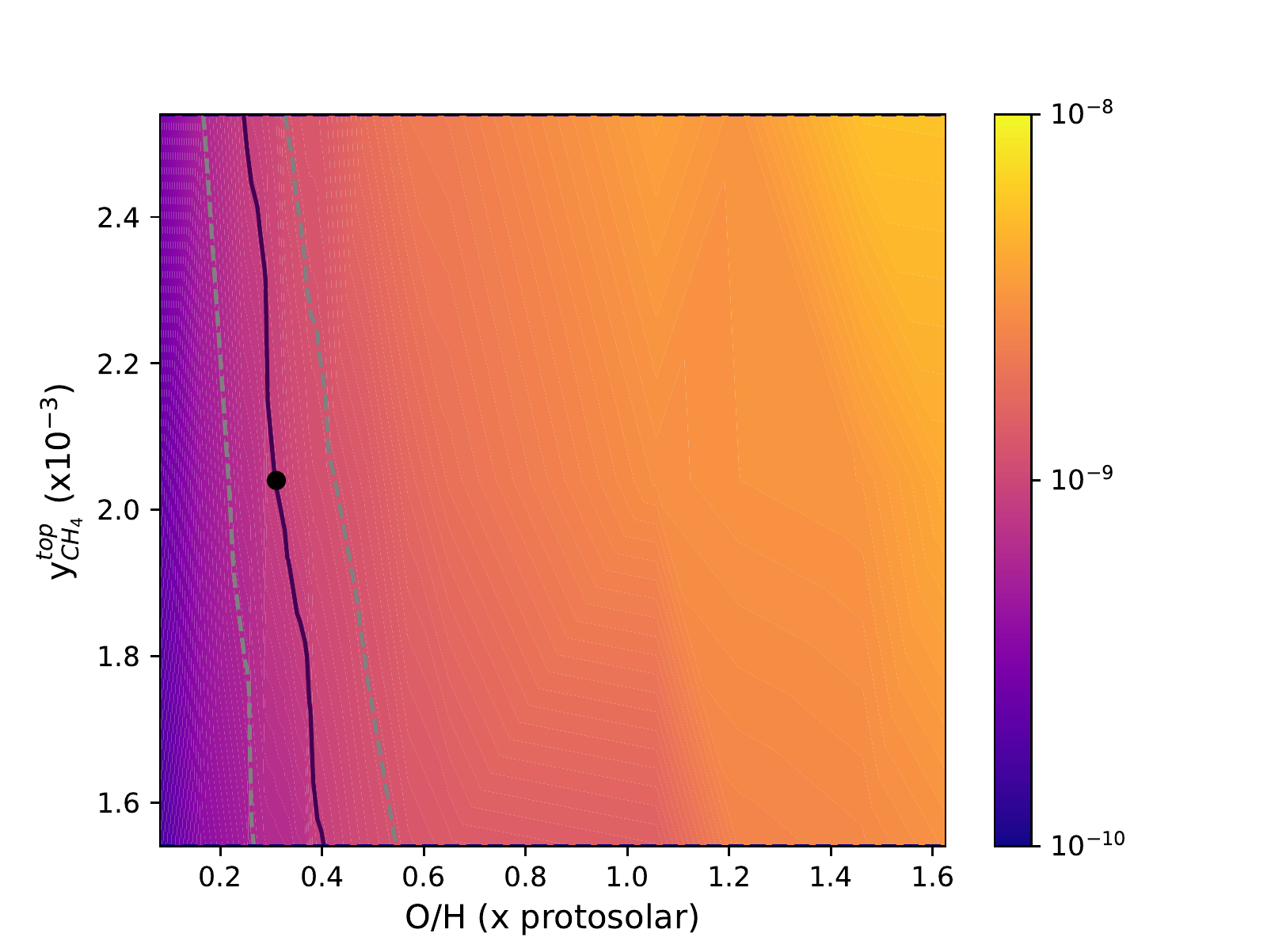}
\end{center}
\caption{Carbon and oxygen dependence of Jupiter's upper tropospheric CO mole fraction. The CO mole fraction (colour scale) as a function of the carbon abundance (represented by $y_\mathrm{CH_4}^\mathrm{top}$) and oxygen water abundance relative to protosolar according to the thermochemical model results when assuming a constant $K_\mathrm{zz}$  of 10$^8$\Kunit. The layout is similar to fig{Fig2}.}
\label{Fig3} 
\end{figure*}

With our nominal parameter set (Methods), we found that a deep oxygen abundance of 0.3 times the protosolar value is required in Jupiter's deep troposphere. The corresponding vertical profiles are displayed in \fig{Fig1}. In our simulations, we allowed three parameters to vary to fit the tropospheric CO and estimate the uncertainty of the deep oxygen abundance with respect to the methane mole fraction $y_\mathrm{CH_4}^\mathrm{top}$ at the top of the troposphere and the vertical eddy diffusion coefficient $K_\mathrm{zz}$. We allowed these two parameters to vary within their respective uncertainty ranges; that is, $y_\mathrm{CH_4}^\mathrm{top}$ = 0.00204 $\pm$ 0.0050 (ref. 3) and $K_\mathrm{zz}$ $=$ 10$^8$\Kunit~within a factor of two \citep{Wang2016,Grassi2020}. When $y_\mathrm{CH_4}^\mathrm{top}$ was allowed to vary within its uncertainty range and $K_\mathrm{zz}$ was fixed to 10$^8$\Kunit, we obtained our nominal fits to the CO mole fraction with a deep oxygen abundance of 0.3$_{-0.2}^{+0.3}$ times the protosolar value (\fig{Fig2}). Conversely, when we varied $K_\mathrm{zz}$ and fixed $y_\mathrm{CH_4}^\mathrm{top}$ to 0.00204, we obtained a deep oxygen abundance of 0.3$_{-0.2}^{+0.3}$ times the protosolar value (\fig{Fig3}). The deep carbon abundance was 3.2 $\pm$ 0.8 times the protosolar value, depending on the adopted value of $y_\mathrm{CH_4}^\mathrm{top}$. Finally, if we accounted for the uncertainty ranges of $K_\mathrm{zz}$ and $y_\mathrm{CH_4}^\mathrm{top}$, then the result for the deep oxygen was 0.3$_{-0.2}^{+0.5}$ times the protosolar value. According to our model, oxygen is subsolar and the C/O ratio is 6$_{-5}^{+10}$, suggesting that Jupiter could have a surprisingly carbon-rich envelope.

The deep oxygen abundance in giant planets has long been debated as it is one of the key elements pertaining to the formation of solids and trapping of more volatile species in the protosolar nebula, which are later released to the growing envelopes of the giant planets \citep{Owen1999,Gautier2001}. Remote sensing observations and in situ measurements provide values ranging from 0.25 to 4.2 times the protosolar value. The Galileo measurement, when translated into an O/H abundance using the protosolar composition of \citet{Lodders2021}, is 0.37 $\pm$ 0.12 \citep{Wong2004}. This value has often been considered a lower limit because the water abundance was still increasing in the measurements when the signal from the Galileo probe was lost, and the probe entered a dry region of Jupiter's atmosphere (a 5 $\mu$m hotspot). \citet{Bjoraker2018} found that their Great Red Spot spectrum around 5 $\mu$m was best fitted by fixing the water cloud base at 5 $\pm$ 1 bar, which translates nominally into a near-protosolar deep oxygen abundance, but the 1$\sigma$  range encompasses 0.3 to 3 times the protosolar value. Finally, a deep water abundance of 2.2$_{-1.4}^{+2.0}$ times the protosolar value was obtained \citep{Li2020} with the Juno Microwave Spectrometer (MWR) in the sole equatorial region where vertical mixing and meteorology seem to maintain a well-mixed atmosphere throughout the probed gas column. Whether this value is representative of the whole planet remains to be verified, especially given the unexpected results for the meridional distribution of ammonia and its depletion at pressures lower than 30 bar \citep{Li2017} and the role water seems to play in this depletion \citep{Guillot2020a}. The fact that \citet{Li2017} found about half the deep ammonia measured by Galileo could indicate that this oxygen measurement is also not representative of the global value. In any case, their oxygen abundance retrieval presents an error bar in which the lower end is more weakly determined than the higher end. The lower 2$\sigma$ limit lies at 0.1 times the protosolar value.

\section{Discussion}
We nominally found a subsolar deep oxygen abundance in Jupiter. Our results, compatible with the range obtained from similar modelling in \citet{Visscher2010}, indicate marginal agreement with the Juno MWR analysis \citep{Li2020} as discussed above. Cloud models often require one times solar oxygen or higher \citep{Inurrigarro2022}, but can also accommodate subsolar oxygen \citep{Hueso2001}. More problematic are the results from lightning data. While overall modelling of lightning frequency \citep{Aglyamov2021} permits a subsolar value, the Galileo detection of a potentially deeper lightning flash would imply a water enrichment exceeding solar values \citep{Dyudina2002}. Decreasing $K_\mathrm{zz}$ from the nominal value of 10$^8$\Kunit~to 2.5 $\times$ 10$^6$\Kunit~would raise the deep oxygen abundance from 0.3 to 2.2 times solar (\fig{Fig1} and Extended Data Fig. 2). However, such a low level of vertical mixing must be confined to the altitudes below which temperatures reach 1000--1100 K, where CO quenching starts, as demonstrated in \fig{Fig1}; at 800 K it must be at or close to our nominal value to fit the data on other disequilibrium species such as PH$_3$ and GeH$_4$ \citep{Wang2016,Grassi2020}, as described in Methods. Our revised chemical network, taken from \citet{Venot2020}, probably still suffers from some uncertainties in the reaction rates. However, any change in the rates to produce a solar or supersolar water abundance would require the deep oxygen abundance derived in the ice giants \citep{Venot2020} to increase as well, raising the problem of the consistency between the deep oxygen abundance and deep D/H ratio when compared with that found in Oort cloud comets \citep{Ali-Dib2014}. Our results therefore require either (1) a carbon-rich envelope in Jupiter or (2) a deep layer of reduced vertical mixing.

Option (1) was already proposed by \citet{Mousis2012}, assuming that the Galileo O determination corresponds to the bulk abundance. A high C/O ratio in Jupiter's envelope such as that of 6$^{+10}_{-5}$ found here resulting from a relatively low oxygen abundance was also proposed by previous studies \citep{Lodders2004,Mousis2019}. In one model \citep{Mousis2019}, the radial drift of pebbles through the amorphous-to-crystalline-ice transition front in the protoplanetary disk releases carbon-rich supervolatiles into the gas while stranding water in the ice. Other scenarios include the agglomeration of building blocks condensed in the vicinity of the condensation lines of C-rich volatiles by the growing Jupiter. The C/O ratio in icy solids formed in those regions is expected to rise steeply, as found by \citet{Mousis2021a} to explain the water-poor composition of Comet C/2016 R2 (PanSTARRS). Another study \citep{Mousis2021b} suggested that a wide range of protosolar nebula compositions can match Jupiter's metallicity, including several types of icy phase (clathrates and pure condensates). It should be noted that current Jupiter formation and structure models predict a low-metallicity envelope \citep{Helled2022,Schneider2021}.

Option (2) implies a dramatic decrease in vertical mixing at temperatures higher than 1000--1100 K, corresponding to a pressure level of roughly 0.6 kbar. Various mechanisms may lead to stable regions throughout Jupiter's deep atmosphere, and one mechanism that corresponds roughly to the p-T region here is a radiative zone extending from 1200--2000 K \citep{Guillot1994}. Such a zone is the result of low opacity, obtained only in the case of a depletion in the alkali metals \citep{Guillot2004}, and Juno MWR data hint at such a depletion \citep{Bhattacharya2021}. It is therefore plausible that the vertical mixing, represented by $K_\mathrm{zz}$ in our model, is very low in the region just below the CO quench level, begins to increase at that level and reaches its full convective value by 900 K (300 bar) where PH$_3$ and GeH$_4$ quenching determine our nominal value for $K_\mathrm{zz}$. A model in which $K_\mathrm{zz}$ was set to the very low value of 1\Kunit~(it could in principle be as low as the molecular diffusivity) at temperatures higher than 1400 K and to 10$^8$\Kunit~at temperatures lower than 970 K and then interpolated logarithmically between the two levels produced satisfactory results with solar oxygen (\fig{Fig1} and Extended Data Fig. 2).

Either of the two possibilities -- depleted oxygen or a deep radiative zone -- would be important for understanding the nature of Jupiter's interior below its visible atmosphere. Further analysis of Juno MWR data during the extended mission, particularly at the longest wavelength channel, will help to distinguish between them. The results for Jupiter also provide important context for the elemental composition of giant planets in our Solar System, and beyond. In situ probes, despite their inherent limitation of a single entry-point, would provide
invaluable compositional data for Saturn and the ice giants, especially regarding noble gases, as presented other works \citep{Mousis2014,Mousis2018}. \citet{Cavalie2020} have shown how the use of thermochemical simulations can increase the science return of in situ probe measurements.

\section*{Methods}
\subsection*{Thermochemical model}
We used the 1D thermochemical and diffusion model initially developed in \citet{Venot2012} for warm exoplanets, and adapted in \citet{Cavalie2014,Cavalie2017} to study the deep oxygen abundance in Uranus and Neptune. The model solves the continuity equation as a function of time at each altitude, for 111 carbon, oxygen and nitrogen species through 1912 reactions. 

Our model required boundary conditions regarding the composition of the upper troposphere. We took the upper tropospheric mole fraction of He and CH$_4$ from \citet{vonZahn1998} and \citet{Wong2004}, respectively $y_\mathrm{He}^\mathrm{top}$ $=$ 0.1359 $\pm$ 0.0027 and $y_\mathrm{CH_4}^\mathrm{top}$ $=$ 0.00204 $\pm$ 0.0050, both resulting from the Galileo probe measurements. For CO, we adopted an upper tropospheric mole fraction of 0.9 $\pm$ 0.3 ppb, following measurements from \citet{Bezard2002} and \citet{Bjoraker2018}.

To constrain the deep oxygen abundance of Jupiter from upper tropospheric CO observations, we also needed to make assumptions on the vertical mixing and on the temperature profile. Following \citet{Wang2015} and recent Juno results of \citet{Grassi2020} on tropospheric abundances of disequilibrium species, we nominally adopted a vertical eddy mixing coefficient of 10$^8$\Kunit~with a factor of two uncertainty. Even though Juno observations of NH$_3$ \citep{Li2017} and models to explain downward ammonia transport \citep{Guillot2020a,Guillot2020b} show that chemical transport does not obey a pure diffusion equation between 0.1 and tens of bars, we assumed that this is the case at several hundred bars in the more homogeneously mixed deeper troposphere where the CO thermochemistry is quenched. Our temperature profile follows the Galileo profile \citep{Seiff1998} within 1 K down to 22 bar and we extrapolated temperatures using a wet adiabat from the diffusion-dominated upper levels down to the deep levels where thermochemical equilibrium prevails. The deep oxygen abundance measured with Juno for Jupiter \citep{Li2020} is not high enough to produce a radiative layer, resulting from a mean molecular weight gradient at the water condensation level in which the temperature would sharply increase, as opposed to the case of the ice giants \citep{Cavalie2017}. We stopped our temperature extrapolations at 1700 K, much deeper than the levels where thermochemistry is quenched by vertical mixing. We ensured that our results reached steady state with an integration time of 10$^{10}$ s.

We fixed the deep nitrogen abundance to $\sim$4 times the protosolar value to reproduce the Galileo measurement \citep{Wong2004}. The abundance of N$_2$ was then $\sim$10$^{-5}$ in the upper troposphere. Our model results regarding oxygen were, however, insensitive to this value, because the nitrogen and oxygen chemistries were mostly uncoupled. 

We adopted the protosolar abundances reported in \citet{Lodders2021} throughout this Article. We used them to express our model results and to convert previous results to a common scale.

\subsection*{Chemical scheme}
The chemical network in \citet{Venot2012} is a C/H/O/N mechanism initially validated for the combustion industry to help to understand the combustion of fuels in car engines and thus limit their environmental impact. It is based on a C$_0$--C$_2$ mechanism to which a nitrogen reaction base was added. It comprises 105 species and 1926 reactions. Although it was validated against experiments for pressures ranging from 0.01 bar to several hundred bars and for temperatures ranging from 300 to 2500 K, \citet{Wang2016} showed that the conversion of H$_2$O into CO was less efficient in Jupiter with the network shown in \citet{Venot2012} compared with several others (see their fig. 17), resulting in CO abundance 10 times lower than in simulations involving competing networks. A similar issue with CH$_4$?CO chemistry had been found in applications to hot Jupiters by \citet{Moses2014}, even though \citet{Venot2020} found even more compelling differences in cooler planets. \citet{Moses2014} further narrowed down the main difference in the networks to the kinetics of methanol through the H $+$ CH$_3$OH $=$ CH$_3$ $+$ H$_2$O reaction. The chemical rate of this reaction had been set in \citet{Venot2012} to that estimated by \citet{Hidaka1989}, but was found to be over-estimated by \citet{Visscher2010} in their work on Jupiter's thermochemistry. This led \citet{Venot2020} to fully revise the CH$_3$OH sub-network of their chemical scheme. They adopted experimental data \citep{Burke2016} that are remarkable in several aspects. First, the reaction of \citet{Hidaka1989} is no longer explicitly present in the network. This does not prevent CH$_3$OH from being destroyed and producing CH$_3$ and H$_2$O, but this is achieved through other destruction pathways. Second, the kinetic rates of several reactions of this network (more specifically those involving methoxide and the methyl radical) have an explicit logarithmic dependence with pressure defined for up to five pressure decades, which thus increase the accuracy and robustness of the kinetics over this wide range of pressure conditions. This new scheme was validated \citep{Venot2020} over a wide range of temperature and pressure conditions and showed improved agreement with experimental data. We have produced a CO profile for Jupiter in the same conditions as in fig. 17 of \citet{Wang2016}. It is shown in Extended Data Fig. 1 and it fully agrees with the profiles presented in fig. 17 of \citet{Wang2016}. When applied to the ice giants \citep{Venot2020} this scheme resulted in the production of the observed CO in ice giants with substantially lower oxygen enrichments.

As stated by \citet{Moses2014}, ``the exact mechanism involved with CH$_4$--CO quenching in reducing environments has not been strictly identified''. This is caused by the high nonlinearity and high coupling between the various chemical reactions of the scheme. It is thus not possible, as found in initial studies that assumed a rate-limiting reaction in the CO destruction mechanism \citep{Fegley1988,Yung1988,Bezard2002}, to easily identify a single reaction and quantify uncertainties on the results from the uncertainty on the rate of this reaction. A methodology of uncertainty propagation and global sensitivity analysis has been developed \citep{Dobrijevic2010}, but it required running several hundred simulations following a Monte Carlo scheme. Applying this methodology is beyond the scope of this study.

\begin{figure*}[!h]
\begin{center}
   \includegraphics[width=15cm,keepaspectratio]{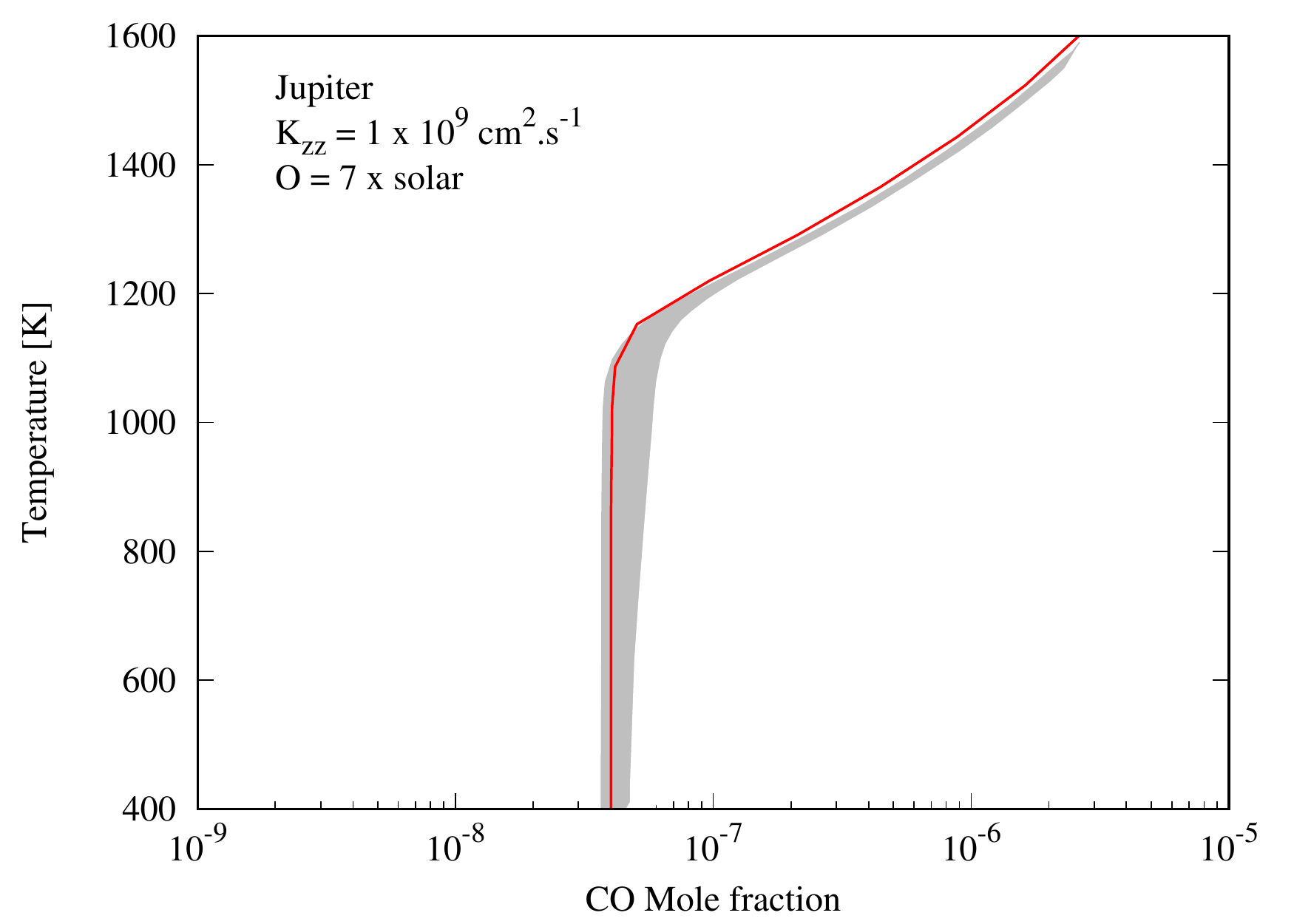}
\end{center}
\caption{CO vertical profile in Jupiter computed in the same conditions as in (15) with our chemical scheme, i.e., that of (21) with revised methanol chemistry kinetics. The profile is obtained for $K_\mathrm{zz}$ $=$ 10$^9$\Kunit~and seven times solar oxygen. It is in full agreement with those obtained with other chemical schemes and shown in Figure 17 of (15), which are indicated by the grey area.}
\label{FigS1} 
\end{figure*}

\subsection*{Deep radiative region in Jupiter?}
It has been shown \citep{Wang2016} that PH$_3$ and GeH$_4$ are quenched at $\sim$700--800 K ($p$ $\approx$ 0.1 kbar; see their figs. 6 and 11) and the abundances observed with Juno JIRAM \citep{Grassi2020} in the upper troposphere imply that $K_\mathrm{zz}$ $\approx$ 10$^7$--10$^9$\Kunit~from PH$_3$ and $\sim$10$^8$\Kunit~from GeH$_4$ at this level. For uniform $K_\mathrm{zz}$ $>$ 10$^7$\Kunit~our model predicts a subsolar oxygen to fit the observed abundance of CO (Extended Data Table 1).

The first hint that Jupiter's troposphere could harbour a radiative region in the vicinity of the layers where CO is quenched ($T$ $\approx$ 1000--1100 K; \fig{Fig1}) was then obtained when the deep oxygen abundance was raised to the Juno MWR nominal value of 2.2 times protosolar. This required us to decrease the vertical mixing $K_\mathrm{zz}$ from its nominal value of 10$^8$\Kunit~to 2.5 $\times$ 10$^6$\Kunit~(\fig{Fig1}). Fitting the whole range of 1$\sigma$ uncertainties of the Juno measurement led to the $K_\mathrm{zz}$ value reported in Extended Data Table 1. We thus found that convection needs to be less efficient in the CO quench region, with $K_\mathrm{zz}$ lowered by a factor 10 to 100, to obtain solar-to-supersolar oxygen.

A decrease in the Rosseland opacity due to hydrogen and helium opacity between 1200 and 4000 K in Jupiter can result in a radiative region, as initially pointed out by a previous study \citep{Guillot1994}. It was subsequently confirmed \citep{Guillot2004} that this region may exist between $p$ $\approx$ 1.5 kbar ($T$ $\approx$ 1400 K) and $p$ $\approx$ 8.0 kbar ($T$ $\approx$ 2200 K) provided that the Jovian atmosphere is also depleted in alkali metals. This depletion seems to be confirmed by recent Juno MWR observations \citep{Bhattacharya2021}. If such a radiative region exists, vertical heat transport and chemical mixing would strongly be inhibited, and our assumption of a vertically uniform $K_\mathrm{zz}$ would not hold in this region. A previous work \citep{Cavalie2017} investigated the effect of an insulation layer produced in the ice giants by the rapid change in mean molecular weight gradient where water condenses. Despite the presence of this insulation layer in the altitudes where CO is quenched and vertical transport prevails, they found very limited impact on their results, mostly because the radiative layer was very thin. The radiative layer in Jupiter, which is of a different origin than that in the ice giants, may extend from 1.5 to 8 kbar \citep{Guillot2004}. Even if the radiative layer itself
has a limited impact on the CO profile, because it is located in the region where thermochemical equilibrium between CO and H$_2$O prevails over vertical transport, we needed to assess the effect of such a layer and how it connects to upper layers in our simulations.

We found that our model could reconcile solar oxygen with the observed tropospheric CO by adding the presence of a deep radiative layer in which $K_\mathrm{zz}$ could be as low as the molecular diffusivity. We set $K_\mathrm{zz}$ to 1\Kunit~for layers with $T$ $>$ 1400 K and interpolated $K_\mathrm{zz}$ between this value at 1400 K and our nominal value of 10$^8$\Kunit~at 970 K, such that PH$_3$ and GeH$_4$ are quenched in the $\sim$800 K region as expected from models and observations. This ensures that $K_\mathrm{zz}$ is low enough where CO is quenched. The resulting CO profile is shown in \fig{Fig1}.

The $K_\mathrm{zz}$ profiles used in this work that correspond to the results presented in \fig{Fig1} are shown in Extended Data Fig. 2.

\begin{table}
\begin{center}
\begin{tabular}{cc}
\hline
O/H ($\times$ the protosolar value) & Required $K_\mathrm{zz}$ in the CO quench region \\
\hline
0.3 & 1$\times$10$^8$\Kunit\\
0.8 & 2.5$\times$10$^7$\Kunit\\
2.2 & 2.5$\times$10$^6$\Kunit\\
4.2 & 4$\times$10$^5$\Kunit\\
\hline
\end{tabular}
\end{center}
\caption{Relationship between Jupiter's deep O/H and the $K_\mathrm{zz}$ required in the quench region of CO to fit the observed upper tropospheric CO mole fraction. This essentially shows that the higher the deep oxygen abundance, the more inhibited the mixing to produce the right abundance of CO.}
\label{TableS1}
\end{table}

\begin{figure*}[!h]
\begin{center}
   \includegraphics[width=15cm,keepaspectratio]{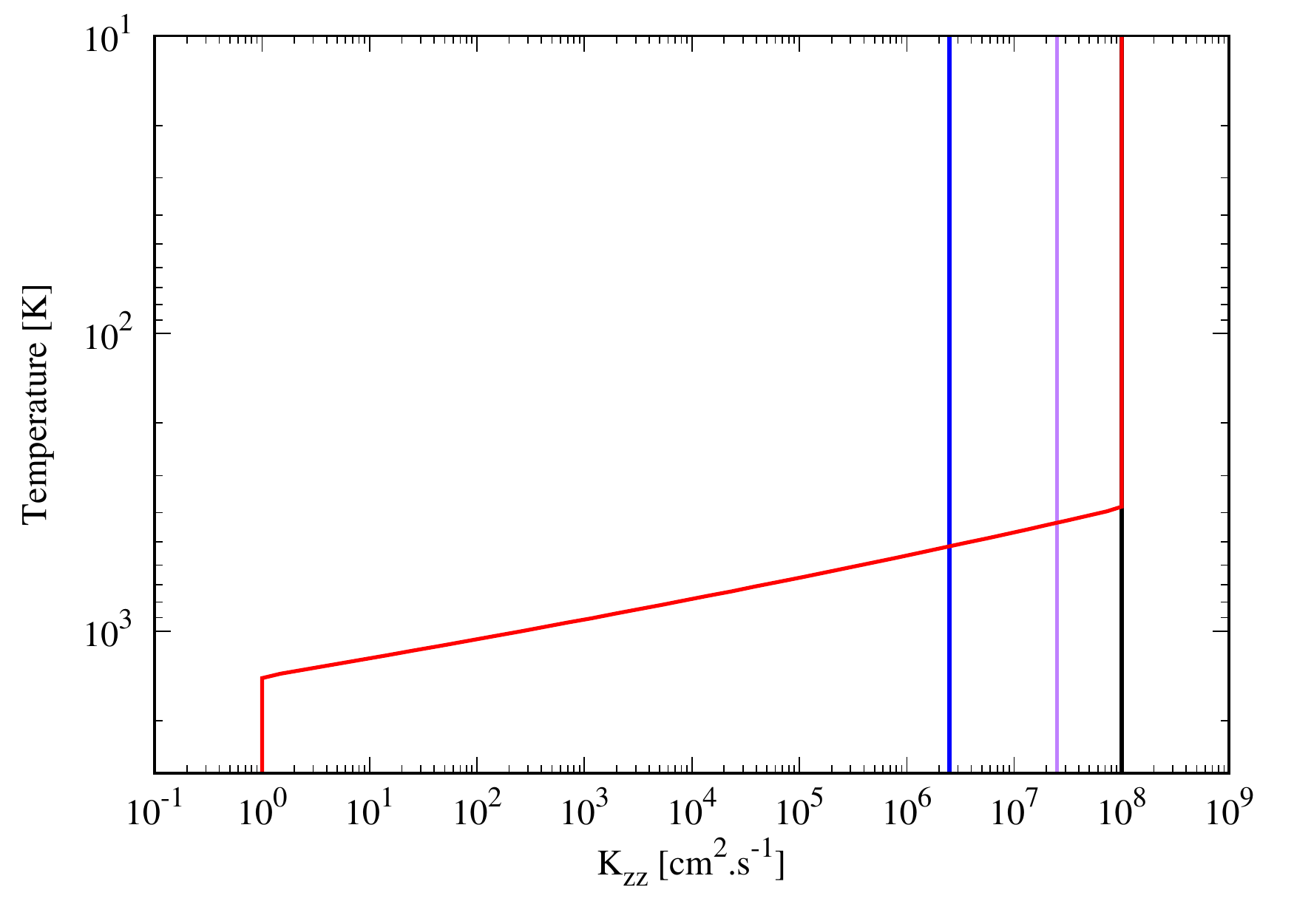}
\end{center}
\caption{$K_\mathrm{zz}$ profiles used in this work. The black profile is our nominal model (where $K_\mathrm{zz}$ $=$10$^8$\Kunit, constant with altitude) which results in an oxygen abundance of 0.3 times the protosolar value. The blue profile ($K_\mathrm{zz}$ $=$2.5 $\times$ 10$^6$\Kunit, constant with altitude) results constrains oxygen to 2.2 times the protosolar value, i.e., the Juno MWR nominal measurement of (7). An intermediate constant value of 2.5 $\times$ 10$^7$\Kunit~(purple line) will produce the observed CO with nearly solar oxygen. The red profile (variable with altitude) indicates the presence of a stable radiative layer at depth with a transition region such that $K_\mathrm{zz}$ reaches our nominal value at the levels where PH$_3$ and GeH$_4$ are quenched.}
\label{FigS2} 
\end{figure*}


\section*{Acknowledgements}
T.C. acknowledges funding from CNES and the Programme National de Plan\'etologie (PNP) of CNRS/INSU. J.L. acknowledges support from the Juno mission through a subcontract from the Southwest Research Institute.


\bibliographystyle{aa}
\bibliography{Jupiter_deep_water}

\begin{thebibliography}{50}
\expandafter\ifx\csname natexlab\endcsname\relax\def\natexlab#1{#1}\fi

\bibitem[{{Aglyamov} {et~al.}(2021){Aglyamov}, {Lunine}, {Becker}, {Guillot},
  {Gibbard}, {Atreya}, {Bolton}, {Levin}, {Brown}, \& {Wong}}]{Aglyamov2021}
{Aglyamov}, Y.~S., {Lunine}, J., {Becker}, H.~N., {et~al.} 2021, Journal of
  Geophysical Research (Planets), 126, e06504

\bibitem[{{Ali-Dib} {et~al.}(2014){Ali-Dib}, {Mousis}, {Petit}, \&
  {Lunine}}]{Ali-Dib2014}
{Ali-Dib}, M., {Mousis}, O., {Petit}, J.-M., \& {Lunine}, J.~I. 2014, \apj,
  793, 9

\bibitem[{{Bar-Nun} {et~al.}(1988){Bar-Nun}, {Kleinfeld}, \&
  {Kochavi}}]{Bar-Nun1988}
{Bar-Nun}, A., {Kleinfeld}, I., \& {Kochavi}, E. 1988, \prb, 38, 7749

\bibitem[{{Beer}(1975)}]{Beer1975}
{Beer}, R. 1975, \apjl, 200, L167

\bibitem[{{B{\'e}zard} {et~al.}(2002){B{\'e}zard}, {Lellouch}, {Strobel},
  {Maillard}, \& {Drossart}}]{Bezard2002}
{B{\'e}zard}, B., {Lellouch}, E., {Strobel}, D., {Maillard}, J.-P., \&
  {Drossart}, P. 2002, \icarus, 159, 95

\bibitem[{{Bhattacharya} {et~al.}(2021){Bhattacharya}, {Li}, {Atreya},
  {Steffes}, {Levin}, \& {Bolton}}]{Bhattacharya2021}
{Bhattacharya}, A., {Li}, C., {Atreya}, S., {et~al.} 2021, in AAS/Division for
  Planetary Sciences Meeting Abstracts, Vol.~53, AAS/Division for Planetary
  Sciences Meeting Abstracts, 212.01

\bibitem[{{Bjoraker} {et~al.}(2018){Bjoraker}, {Wong}, {de Pater}, {Hewagama},
  {{\'A}d{\'a}mkovics}, \& {Orton}}]{Bjoraker2018}
{Bjoraker}, G.~L., {Wong}, M.~H., {de Pater}, I., {et~al.} 2018, \aj, 156, 101

\bibitem[{{Burke} {et~al.}(2016){Burke}, {Metcalfe}, {Burke}, {Heufer},
  {Dagaut}, \& {Curran}}]{Burke2016}
{Burke}, U., {Metcalfe}, W.~K., {Burke}, S.~M., {et~al.} 2016, \combust, 165,
  125

\bibitem[{{Cavali{\'e}} {et~al.}(2014){Cavali{\'e}}, {Moreno}, {Lellouch},
  {Hartogh}, {Venot}, {Orton}, {Jarchow}, {Encrenaz}, {Selsis}, {Hersant}, \&
  {Fletcher}}]{Cavalie2014}
{Cavali{\'e}}, T., {Moreno}, R., {Lellouch}, E., {et~al.} 2014, \aap, 562, A33

\bibitem[{{Cavali{\'e}} {et~al.}(2020){Cavali{\'e}}, {Venot}, {Miguel},
  {Fletcher}, {Wurz}, {Mousis}, {Bounaceur}, {Hue}, {Leconte}, \&
  {Dobrijevic}}]{Cavalie2020}
{Cavali{\'e}}, T., {Venot}, O., {Miguel}, Y., {et~al.} 2020, \ssr, 216, 58

\bibitem[{{Cavali{\'e}} {et~al.}(2017){Cavali{\'e}}, {Venot}, {Selsis},
  {Hersant}, {Hartogh}, \& {Leconte}}]{Cavalie2017}
{Cavali{\'e}}, T., {Venot}, O., {Selsis}, F., {et~al.} 2017, \icarus, 291, 1

\bibitem[{{de Pater}(1986)}]{dePater1986}
{de Pater}, I. 1986, \icarus, 68, 344

\bibitem[{{Dobrijevic} {et~al.}(2010){Dobrijevic}, {Cavali{\'e}},
  {H{\'e}brard}, {Billebaud}, {Hersant}, \& {Selsis}}]{Dobrijevic2010}
{Dobrijevic}, M., {Cavali{\'e}}, T., {H{\'e}brard}, E., {et~al.} 2010, \planss,
  58, 1555

\bibitem[{{Dyudina} {et~al.}(2002){Dyudina}, {Ingersoll}, {Vasavada}, {Ewald},
  \& {Galileo SSI Team}}]{Dyudina2002}
{Dyudina}, U.~A., {Ingersoll}, A.~P., {Vasavada}, A.~R., {Ewald}, S.~P., \&
  {Galileo SSI Team}. 2002, \icarus, 160, 336

\bibitem[{{Fegley} \& {Prinn}(1988)}]{Fegley1988}
{Fegley}, B. \& {Prinn}, R.~G. 1988, \apj, 324, 621

\bibitem[{{Gautier} {et~al.}(2001){Gautier}, {Hersant}, {Mousis}, \&
  {Lunine}}]{Gautier2001}
{Gautier}, D., {Hersant}, F., {Mousis}, O., \& {Lunine}, J.~I. 2001, \apjl,
  550, L227

\bibitem[{{Grassi} {et~al.}(2020){Grassi}, {Adriani}, {Mura}, {Atreya},
  {Fletcher}, {Lunine}, {Orton}, {Bolton}, {Plainaki}, {Sindoni}, {Altieri},
  {Cicchetti}, {Dinelli}, {Filacchione}, {Migliorini}, {Moriconi}, {Noschese},
  {Olivieri}, {Piccioni}, {Sordini}, {Stefani}, {Tosi}, \&
  {Turrini}}]{Grassi2020}
{Grassi}, D., {Adriani}, A., {Mura}, A., {et~al.} 2020, Journal of Geophysical
  Research (Planets), 125, e06206

\bibitem[{{Guillot} {et~al.}(1994){Guillot}, {Gautier}, {Chabrier}, \&
  {Mosser}}]{Guillot1994}
{Guillot}, T., {Gautier}, D., {Chabrier}, G., \& {Mosser}, B. 1994, \icarus,
  112, 337

\bibitem[{{Guillot} {et~al.}(2020{\natexlab{a}}){Guillot}, {Li}, {Bolton},
  {Brown}, {Ingersoll}, {Janssen}, {Levin}, {Lunine}, {Orton}, {Steffes}, \&
  {Stevenson}}]{Guillot2020b}
{Guillot}, T., {Li}, C., {Bolton}, S.~J., {et~al.} 2020{\natexlab{a}}, Journal
  of Geophysical Research (Planets), 125, e06404

\bibitem[{{Guillot} {et~al.}(2020{\natexlab{b}}){Guillot}, {Stevenson},
  {Atreya}, {Bolton}, \& {Becker}}]{Guillot2020a}
{Guillot}, T., {Stevenson}, D.~J., {Atreya}, S.~K., {Bolton}, S.~J., \&
  {Becker}, H.~N. 2020{\natexlab{b}}, Journal of Geophysical Research
  (Planets), 125, e06403

\bibitem[{{Guillot} {et~al.}(2004){Guillot}, {Stevenson}, {Hubbard}, \&
  {Saumon}}]{Guillot2004}
{Guillot}, T., {Stevenson}, D.~J., {Hubbard}, W.~B., \& {Saumon}, D. 2004, in
  Jupiter. The Planet, Satellites and Magnetosphere, ed. F.~{Bagenal}, T.~E.
  {Dowling}, \& W.~B. {McKinnon}, Vol.~1, 35--57

\bibitem[{{Helled} \& {Lunine}(2014)}]{Helled2014}
{Helled}, R. \& {Lunine}, J. 2014, \mnras, 441, 2273

\bibitem[{{Helled} {et~al.}(2022){Helled}, {Stevenson}, {Lunine}, {Bolton},
  {Nettelmann}, {Atreya}, {Guillot}, {Militzer}, {Miguel}, \&
  {Hubbard}}]{Helled2022}
{Helled}, R., {Stevenson}, D.~J., {Lunine}, J.~I., {et~al.} 2022, \icarus, 378,
  114937

\bibitem[{{Hidaka} {et~al.}(1989){Hidaka}, {Oki}, {Kawano}, \&
  {Higashihara}}]{Hidaka1989}
{Hidaka}, Y., {Oki}, T., {Kawano}, H., \& {Higashihara}, T. 1989, \jchemphys,
  93, 7134

\bibitem[{{Hueso} \& {S{\'a}nchez-Lavega}(2001)}]{Hueso2001}
{Hueso}, R. \& {S{\'a}nchez-Lavega}, A. 2001, \icarus, 151, 257

\bibitem[{{I{\~n}urrigarro} {et~al.}(2022){I{\~n}urrigarro}, {Hueso},
  {S{\'a}nchez-Lavega}, \& {Legarreta}}]{Inurrigarro2022}
{I{\~n}urrigarro}, P., {Hueso}, R., {S{\'a}nchez-Lavega}, A., \& {Legarreta},
  J. 2022, \icarus, 386, 115169

\bibitem[{{Janssen} {et~al.}(2005){Janssen}, {Hofstadter}, {Gulkis},
  {Ingersoll}, {Allison}, {Bolton}, {Levin}, \& {Kamp}}]{Janssen2005}
{Janssen}, M.~A., {Hofstadter}, M.~D., {Gulkis}, S., {et~al.} 2005, \icarus,
  173, 447

\bibitem[{{Li} {et~al.}(2020){Li}, {Ingersoll}, {Bolton}, {Levin}, {Janssen},
  {Atreya}, {Lunine}, {Steffes}, {Brown}, {Guillot}, {Allison}, {Arballo},
  {Bellotti}, {Adumitroaie}, {Gulkis}, {Hodges}, {Li}, {Misra}, {Orton},
  {Oyafuso}, {Santos-Costa}, {Waite}, \& {Zhang}}]{Li2020}
{Li}, C., {Ingersoll}, A., {Bolton}, S., {et~al.} 2020, Nature Astronomy, 4,
  609

\bibitem[{{Li} {et~al.}(2017){Li}, {Ingersoll}, {Janssen}, {Levin}, {Bolton},
  {Adumitroaie}, {Allison}, {Arballo}, {Bellotti}, {Brown}, {Ewald}, {Jewell},
  {Misra}, {Orton}, {Oyafuso}, {Steffes}, \& {Williamson}}]{Li2017}
{Li}, C., {Ingersoll}, A., {Janssen}, M., {et~al.} 2017, \grl, 44, 5317

\bibitem[{{Lodders}(2004)}]{Lodders2004}
{Lodders}, K. 2004, \apj, 611, 587

\bibitem[{{Lodders}(2021)}]{Lodders2021}
{Lodders}, K. 2021, \ssr, 217, 44

\bibitem[{{Lunine} \& {Stevenson}(1987)}]{Lunine1987}
{Lunine}, J.~I. \& {Stevenson}, D.~J. 1987, \icarus, 70, 61

\bibitem[{{Moses}(2014)}]{Moses2014}
{Moses}, J.~I. 2014, Philosophical Transactions of the Royal Society of London
  Series A, 372, 20130073

\bibitem[{{Mousis} {et~al.}(2021{\natexlab{a}}){Mousis}, {Aguichine},
  {Bouquet}, {Lunine}, {Danger}, {Mandt}, \& {Luspay-Kuti}}]{Mousis2021a}
{Mousis}, O., {Aguichine}, A., {Bouquet}, A., {et~al.} 2021{\natexlab{a}},
  \psj, 2, 72

\bibitem[{{Mousis} {et~al.}(2018){Mousis}, {Atkinson}, {Cavali{\'e}},
  {Fletcher}, {Amato}, {Aslam}, {Ferri}, {Renard}, {Spilker}, {Venkatapathy},
  {Wurz}, {Aplin}, {Coustenis}, {Deleuil}, {Dobrijevic}, {Fouchet}, {Guillot},
  {Hartogh}, {Hewagama}, {Hofstadter}, {Hue}, {Hueso}, {Lebreton}, {Lellouch},
  {Moses}, {Orton}, {Pearl}, {S{\'a}nchez-Lavega}, {Simon}, {Venot}, {Waite},
  {Achterberg}, {Atreya}, {Billebaud}, {Blanc}, {Borget}, {Brugger}, {Charnoz},
  {Chiavassa}, {Cottini}, {d'Hendecourt}, {Danger}, {Encrenaz}, {Gorius},
  {Jorda}, {Marty}, {Moreno}, {Morse}, {Nixon}, {Reh}, {Ronnet}, {Schmider},
  {Sheridan}, {Sotin}, {Vernazza}, \& {Villanueva}}]{Mousis2018}
{Mousis}, O., {Atkinson}, D.~H., {Cavali{\'e}}, T., {et~al.} 2018, \planss,
  155, 12

\bibitem[{{Mousis} {et~al.}(2014){Mousis}, {Fletcher}, {Lebreton}, {Wurz},
  {Cavali{\'e}}, {Coustenis}, {Courtin}, {Gautier}, {Helled}, {Irwin}, {Morse},
  {Nettelmann}, {Marty}, {Rousselot}, {Venot}, {Atkinson}, {Waite}, {Reh},
  {Simon}, {Atreya}, {Andr{\'e}}, {Blanc}, {Daglis}, {Fischer}, {Geppert},
  {Guillot}, {Hedman}, {Hueso}, {Lellouch}, {Lunine}, {Murray}, {O`Donoghue},
  {Rengel}, {S{\'a}nchez-Lavega}, {Schmider}, {Spiga}, {Spilker}, {Petit},
  {Tiscareno}, {Ali-Dib}, {Altwegg}, {Bolton}, {Bouquet}, {Briois}, {Fouchet},
  {Guerlet}, {Kostiuk}, {Lebleu}, {Moreno}, {Orton}, \& {Poncy}}]{Mousis2014}
{Mousis}, O., {Fletcher}, L.~N., {Lebreton}, J.-P., {et~al.} 2014, \planss,
  104, 29

\bibitem[{{Mousis} {et~al.}(2021{\natexlab{b}}){Mousis}, {Lunine}, \&
  {Aguichine}}]{Mousis2021b}
{Mousis}, O., {Lunine}, J.~I., \& {Aguichine}, A. 2021{\natexlab{b}}, \apjl,
  918, L23

\bibitem[{{Mousis} {et~al.}(2012){Mousis}, {Lunine}, {Madhusudhan}, \&
  {Johnson}}]{Mousis2012}
{Mousis}, O., {Lunine}, J.~I., {Madhusudhan}, N., \& {Johnson}, T.~V. 2012,
  \apjl, 751, L7

\bibitem[{{Mousis} {et~al.}(2019){Mousis}, {Ronnet}, \& {Lunine}}]{Mousis2019}
{Mousis}, O., {Ronnet}, T., \& {Lunine}, J.~I. 2019, \apj, 875, 9

\bibitem[{{Owen} {et~al.}(1999){Owen}, {Mahaffy}, {Niemann}, {Atreya},
  {Donahue}, {Bar-Nun}, \& {de Pater}}]{Owen1999}
{Owen}, T., {Mahaffy}, P., {Niemann}, H.~B., {et~al.} 1999, \nat, 402, 269

\bibitem[{{Schneider} \& {Bitsch}(2021)}]{Schneider2021}
{Schneider}, A.~D. \& {Bitsch}, B. 2021, \aap, 654, A72

\bibitem[{{Seiff} {et~al.}(1998){Seiff}, {Kirk}, {Knight}, {Young}, {Mihalov},
  {Young}, {Milos}, {Schubert}, {Blanchard}, \& {Atkinson}}]{Seiff1998}
{Seiff}, A., {Kirk}, D.~B., {Knight}, T.~C.~D., {et~al.} 1998, \jgr, 103, 22857

\bibitem[{{Venot} {et~al.}(2020){Venot}, {Cavali{\'e}}, {Bounaceur},
  {Tremblin}, {Brouillard}, \& {Lhoussaine Ben Brahim}}]{Venot2020}
{Venot}, O., {Cavali{\'e}}, T., {Bounaceur}, R., {et~al.} 2020, \aap, 634, A78

\bibitem[{{Venot} {et~al.}(2012){Venot}, {H{\'e}brard}, {Ag{\'u}ndez},
  {Dobrijevic}, {Selsis}, {Hersant}, {Iro}, \& {Bounaceur}}]{Venot2012}
{Venot}, O., {H{\'e}brard}, E., {Ag{\'u}ndez}, M., {et~al.} 2012, \aap, 546,
  A43

\bibitem[{{Visscher} {et~al.}(2010){Visscher}, {Moses}, \&
  {Saslow}}]{Visscher2010}
{Visscher}, C., {Moses}, J.~I., \& {Saslow}, S.~A. 2010, \icarus, 209, 602

\bibitem[{{von Zahn} {et~al.}(1998){von Zahn}, {Hunten}, \&
  {Lehmacher}}]{vonZahn1998}
{von Zahn}, U., {Hunten}, D.~M., \& {Lehmacher}, G. 1998, \jgr, 103, 22815

\bibitem[{{Wang} {et~al.}(2015){Wang}, {Gierasch}, {Lunine}, \&
  {Mousis}}]{Wang2015}
{Wang}, D., {Gierasch}, P.~J., {Lunine}, J.~I., \& {Mousis}, O. 2015, \icarus,
  250, 154

\bibitem[{{Wang} {et~al.}(2016){Wang}, {Lunine}, \& {Mousis}}]{Wang2016}
{Wang}, D., {Lunine}, J.~I., \& {Mousis}, O. 2016, \icarus, 276, 21

\bibitem[{{Wong} {et~al.}(2004){Wong}, {Mahaffy}, {Atreya}, {Niemann}, \&
  {Owen}}]{Wong2004}
{Wong}, M.~H., {Mahaffy}, P.~R., {Atreya}, S.~K., {Niemann}, H.~B., \& {Owen},
  T.~C. 2004, \icarus, 171, 153

\bibitem[{{Yung} {et~al.}(1988){Yung}, {Drew}, {Pinto}, \& {Friedl}}]{Yung1988}
{Yung}, Y.~L., {Drew}, W.~A., {Pinto}, J.~P., \& {Friedl}, R.~R. 1988, \icarus,
  73, 516

\end{thebibliography}

\end{document}